\documentclass[%
 reprint,
 amsmath,amssymb,
 aps,
floatfix,
]{revtex4-2}

\usepackage{graphicx}
\usepackage{dcolumn}
\usepackage{bm}

\begin{document}

\preprint{}

\title{Parisi-Sourlas Supertranslation and Scale without Conformal symmetry}

\author{Yu Nakayama}
\affiliation{%
Yukawa Institute for Theoretical Physics, Kyoto University
}%




\date{\today}

\begin{abstract}
Inspired by the possibility of emergent supersymmetry in critical random systems, we study a field theory model with a quartic potential of one superfield, possessing the Parisi-Sourlas supertranslation symmetry. Within perturbative $\epsilon$ expansion, we find nine non-trivial scale invariant renormalization group fixed points, but only one of them is conformal. We, however, believe scale invariance without conformal invariance cannot occur without a sophisticated mechanism because it predicts the existence of a non-conserved but non-renormalized vector operator called virial current, whose existence must be non-generic.
We show that the virial current in this model is related to the supercurrent by supertranslation. The supertranslation Ward-Takahashi identity circumvents the genericity argument, explaining its non-renormalization property.
\end{abstract}

\maketitle


\section{Introduction}
In equilibrium statistical mechanics, the very symmetry that governs the second-order phase transition and the criticality is not mere scale invariance but conformal invariance. The conformal invariance enables us to use the conformal bootstrap to compute critical exponents of various equilibrium models such as the three-dimensional Ising model with unprecedented accuracy \cite{Kos:2016ysd} (see e.g. \cite{Poland:2018epd,Rychkov:2023wsd} for reviews). The numerical algorithm employed by the conformal bootstrap relies on the reflection positivity in a crucial manner. Coincidentally, it is widely believed that the reflection positivity plays an essential role in the statistical models to show conformal invariance at the critical point as demonstrated most convincingly in two dimensions \cite{Polchinski:1987dy} (and in four dimensions \cite{Luty:2012ww,Dymarsky:2013pqa,Dymarsky:2014zja}). See e.g. \cite{Nakayama:2013is} for a review.

In random statistical systems such as the random field Ising model \cite{Imry:1975zz,Aharony:1976jx} (see e.g. \cite{2018JSP...172..665F,Rychkov:2023rgq} for reviews), the usefulness of conformal symmetry appears less eminent. It is partly because the lack of reflection positivity makes it difficult to use the conformal bootstrap as a semi-definite program. The lack of reflection positivity, however, begs the further question: is the renormalization group fixed point of a random system conformal invariant?

From the group theory viewpoint alone, since scale symmetry is a subgroup of conformal symmetry, we expect much more realizations of scale symmetry than conformal symmetry. In field theories, however, with rotational invariance, there is a severe constraint on the stress tensor to realize scale invariance without conformal invariance. It requires the existence of the vector operator whose divergence is the trace of the stress tensor. It is not conserved but not renormalized either, which should sound non-generic. To ensure such existence, we need a sophisticated mechanism in any interacting field theories.

The only known sophisticated mechanism to resolve the genericity argument so far has been the shift symmetry \cite{Gimenez-Grau:2023lpz}. The shift symmetry relates the operators with different numbers of ``fundamental fields", and in theories with shift symmetry, there exists a natural candidate of the virial current whose scaling dimension is protected thanks to the shift Ward-Takahashi identity. All the field theory models with scale invariance without conformal invariance so far found in the literature \cite{Riva:2005gd,El-Showk:2011xbs,Nakayama:2016cyh,Mauri:2021ili}  had shift symmetries (as far as the authors of \cite{Gimenez-Grau:2023lpz} know).

The goal of this paper is to propose a new mechanism to protect the scaling dimensions of virial current. Physically, our model is inspired by the emergent supersymmetry in critical random systems \cite{Parisi:1979ka,Parisi:1982ud,Kaviraj:2019tbg,Kaviraj:2020pwv,Kaviraj:2021qii}. We study a field theory model with a quartic potential of one superfield, possessing the Parisi-Sourlas supertranslation symmetry. Within perturbative $\epsilon$ expansion, we find nine non-trivial scale invariant renormalization group fixed points, but only one of them is conformal. The novel sophisticated mechanism to protect the scaling dimensions of the virial current in this model is the Parisi-Sourlas supertranslation symmetry.

\section{Fine-tuned Parisi-Sourlas model with supertranslation}
To motivate the emergent supersymmetry in random systems, Parisi and Sourlas started with the {\it classical} stochastic field equations of motion under the random source $h$ \cite{Parisi:1979ka,APYoung_1977}:
\begin{align}
-\partial_\mu^2 \varphi + V'(\varphi) = h \label{classical}
\end{align}
Then they considered the path integral over $\varphi$ that satisfies the classical stochastic equation. The constraint \eqref{classical} can be implemented by the introduction of the Lagrange multiplier field $\omega$:
\begin{align}
Z &= \int  \mathcal{D} \omega \mathcal{D}\varphi \mathcal{D}h  e^{\int d^d x \omega (-\partial_\mu^2 \varphi + V'(\varphi)- h) + R(h)} J[\varphi] \cr
J[\varphi] &=|\mathrm{Det} (-\partial_\mu^2 + V''(\varphi))|  
\end{align}
Here, the Jacobian $J[\varphi]$ ensures the desired constraint, and $\mathcal{D}h e^{R(h)}$ is the random distribution of the source.

 Discarding the absolute value, they further rewrote the Jacobian $J[\varphi]$  by introducing two spinless fermionic fields $\Psi$ and $\bar{\Psi}$ as a path integral:
\begin{align}
Z_{PS} &= \int  \mathcal{D} \omega \mathcal{D}\varphi  \mathcal{D}\Psi \mathcal{D}\bar{\Psi} e^{S_{PS}[\omega,\varphi,\Psi,\bar{\Psi}]} \cr
S_{PS}[\omega,\varphi,\Psi,\bar{\Psi}] = & \int d^d x \omega (-\partial_\mu^2 \varphi + V'(\varphi)) + U(\omega) \cr  & - \Psi(-\partial^2_\mu + V''(\varphi)) \bar{\Psi}
\end{align}
Here, we have replaced $R(h)$ with the corresponding $U(\omega)$ by integrating out $h$.

Although it is not obvious from the construction, we realize that $S_{PS}[\omega,\varphi,\Psi,\bar{\Psi}]$ is invariant under the supertranslation. To see this more explicitly, we introduce the superspace $(x^\mu, \theta, \bar{\theta})$ and the superfield $\Phi(x^\mu,\theta,\bar{\theta}) = \varphi(x^\mu) + \theta \Psi(x^\mu) + \bar{\theta} \bar{\Psi}(x^\mu) + \theta \bar{\theta} \omega(x^\mu)$. Then the Parisi-Sourlas action $S_{PS}$ can be expressed as the superspace integration
\begin{align}
S_{PS} = \int d^dx d^2\theta (\frac{-1}{2}(\Phi\partial_\mu^2\Phi) + \Phi\tilde{U}(\partial_{\bar{\theta}} \partial_{{\theta}}  \Phi) + V(\Phi)).
\end{align}
Here $\omega \tilde{U}(\omega) = U(\omega)$.
The invariance under the supertranslation $\Phi(x^\mu, \theta, \bar{\theta}) \to \Phi(x^\mu, \theta + \epsilon, \bar{\theta} + \bar{\epsilon})$ is manifest. Note that for generic $U$, we do not have the invariance under the superrotation $OSp(d|2)$. In the simplest situation where $R$ and $U$ are quadratic, we do have the superrotation as a symmetry.

In this paper, we will fine-tune $V$ and $U$ (or $R(h)$ originally) so that the classical dimensions of $\omega$, $\varphi$, $\Psi$ and $\bar{\Psi}$ are all equal (i.e. $ \Delta =\frac{d-2}{2}$). In particular, we focus on the bare action with only quartic interactions:
\begin{align}
S =& \int d^dx d^2\theta (\frac{-1}{2}\Phi\partial_\mu^2 \Phi + \lambda_{-2}\Phi(\partial_{\bar{\theta}} \partial_{{\theta}} \Phi)^3  +\frac{\lambda_{1}}{4}\Phi^4  \cr &+ \frac{\lambda_0}{3} \Phi^3 \partial_{\bar{\theta}} \partial_{{\theta}} \Phi  + \frac{\lambda_{-1}}{2} \Phi^2 (\partial_{\bar{\theta}} \partial_{{\theta}} \Phi)^2   ) \cr
=& \int d^d x (-\omega \partial_\mu^2 \varphi + \Psi \partial_\mu^2 \bar{\Psi} + \lambda_{-2} \omega^4 + \lambda_1(\omega \varphi^3 -3 \Psi \bar{\Psi} \varphi^2) \cr
&+ \lambda_0 (\omega^2 \varphi^2 - 2 \Psi\bar{\Psi}\omega \varphi) + \lambda_{-1}(\omega^3 \varphi -  \Psi \bar{\Psi} \omega^2)) .
\end{align}
The upper critical dimension of the model is $d=4$ (rather than $d=6$ when $U$ is quadratic). Here, we have added all the terms (e.g. $\lambda_0$ and $\lambda_{-1}$ terms) that are compatible with the supertranslation, which will be generated under the renormalization group flow. We could have introduced $\omega \partial_\mu^2 \omega = \int d^2 \theta \Phi(\partial_\mu^2 \partial_{\bar{\theta}} \partial_{{\theta}} \Phi)$, but we removed it from the field redefinition of $\varphi \to \varphi + \alpha \omega$.  This action is not invariant under the superrotation: it is invariant only under the separate rotation of $x^\mu$s and $\theta$-$\bar{\theta}$.

\section{Search for scale-invariant fixed points}\label{search}
We now want to find renormalization group fixed points of the action to understand the critical behavior. The one-loop beta function in $d=4-\epsilon$ dimension is universal and can be computed by any convenient method.
\begin{align}
\beta_{\lambda_1} &= -\epsilon \lambda_1 + 24 \lambda_0 \lambda_1  \cr
\beta_{\lambda_0} & = -\epsilon \lambda_0 + 48 \lambda_1 \lambda_{-1} + 16 \lambda_0^2 \cr
\beta_{\lambda_{-1}} & = -\epsilon \lambda_{-1} + 72 \lambda_{-2} \lambda_1 + 32 \lambda_0 \lambda_{-1} \cr
\beta_{\lambda_{-2}} & = -\epsilon \lambda_{-2} + 24 \lambda_0 \lambda_{-2} + 8  \lambda_{-1}^2 
\end{align}
Within this order, the beta functions are related to the trace of the stress tensor as $T^\mu_\mu = \sum_a \beta_a O^a$ up to improvement terms of $\partial^2_\mu (\omega \varphi)$. (At higher order, more careful analysis is needed.)

Before proceeding, let us note that by a $SO(1,1)$ field redefinition $(\omega, \varphi) \to (\omega e^\rho, \varphi e^{-\rho})$ that preserves the kinetic term, the coupling constants transform as $(\lambda_1,\lambda_0,\lambda_{-1},\lambda_{-2}) \to (e^{2\rho}\lambda_1,\lambda_0,e^{-2\rho} \lambda_{-1},e^{-4\rho} \lambda_{-2}) $, and the theories with the coupling constants related by this transformation are physically equivalent. Correspondingly, we hereafter refer to the subscript of $\ \lambda$s as $SO(1,1)$ charge.

Now let us find the fixed points of the renormalization group flow. Demanding $\beta_a = 0$ gives a {\it conformal} fixed point because $T^{\mu}_\mu = 0$. We find that there is (only) one non-trivial conformal fixed point located at $(\lambda_1, \lambda_0,\lambda_{-1},\lambda_{-2}) =(0, \frac{\epsilon}{16}, 0,0)$. This conformal fixed point is stable up to fine-tuning more relevant terms $\Phi^2$ and $\Phi (\partial_{\bar{\theta}} \partial_{{\theta}}  \Phi)$ (or $\varphi \omega - \Psi \bar{\Psi}$ and $\omega^2$).

How do we find scale-invariant but not conformal fixed points? The scale invariance demands that the trace of the stress tensor is given by the divergence of (non-conserved) virial current $V_\mu$: $T^\mu_\mu = \sum_a \beta_a O^a = \partial^\mu V_\mu$. In this model, the natural candidate of the virial current is $V_\mu = \gamma (\omega \partial_\mu \varphi - \varphi \partial_\mu \omega)$. The parameter $\gamma$ is interpreted as an extra degree of freedom in how we introduce the wavefunction renormalization under the change of the scale. (As a technical comment, our one-loop beta function $\beta_a$ here can be identified with an ambiguity-free $B_a$ function in the literature \cite{Fortin:2012hn,Nakayama:2013is}. See Appendix B for more details.)

By using the equations of motion, which is valid at the first order analysis here, the scale invariance demands
\begin{align}
\beta_{\lambda_1} &= -\epsilon \lambda_1 + 24 \lambda_0 \lambda_1 = 2\gamma \lambda_1, \cr
\beta_{\lambda_0} & = -\epsilon \lambda_0 + 48 \lambda_1 \lambda_{-1} + 16 \lambda_0^2 = 0, \cr
\beta_{\lambda_{-1}} & = -\epsilon \lambda_{-1} + 72 \lambda_{-2} \lambda_1 + 32 \lambda_0 \lambda_{-1} = -2 \gamma \lambda_{-1}, \cr
\beta_{\lambda_{-2}} & = -\epsilon \lambda_{-2} + 24 \lambda_0 \lambda_{-2} + 8  \lambda_{-1}^2 = -4 \gamma \lambda_{-2}. \label{scaleinv}
\end{align}
Intuitively, if the coupling constants run in a direction that can be compensated by the field redefinition, the physics does not change. Then we can declare its scale invariance. 

Clearly, the condition \eqref{scaleinv} is a weaker condition than demanding $\beta_a = 0$ because we have extra free parameter $\gamma$. With a free parameter at hand, can we find more fixed points? (We do have a similar free parameter whenever we study perturbative search of scale-invariant fixed points in any scalar field theories within $\epsilon$ expansions, but the expectations that we can find more fixed points this way have always failed(!) if we impose the reflection positivity and compactness. See e.g. \cite{Polchinski:1987dy,Dorigoni:2009ra,Fortin:2012hn,Nakayama:2020bsx,Itsios:2021eig,Papadopoulos:2024uvi,Papadopoulos:2024tgs}.) 

Using Mathematica, in addition to the conformal fixed point with $\gamma=0$ discussed above, we have found eight scale-invariant but not conformal fixed points! They are listed in Table \ref{tab:table1}. All these scale invariant but not conformal fixed points come with a one-parameter family because the fixed points related to the $O(1,1)$ field redefinition are all equivalent. (This is similar to the idea that if the virial current generates the field redefinition of the compact group, it would give the cyclic behavior in the renormalization group flow \cite{Fortin:2012hn}.)
\begin{table}[b]%
\caption{\label{tab:table1}%
Non-trivial scale invariant fixed points with supertranslation symmetry that we found. In the table, we have fixed the $SO(1,1)$ field redefinition ambiguity by $\lambda_{-1} = \epsilon$ (or $\lambda_1 = \epsilon$ or $\lambda_{-2} = \epsilon$ if they are zero). Instability means the number of unstable renormalization group directions.
}
\begin{ruledtabular}
\begin{tabular}{lcccccc}
\textrm{Label}&
$\lambda_1$ &
$\lambda_{0}$&
$\lambda_{-1}$&
$\lambda_{-2}$&
$\gamma$&
instability  \cr
\colrule
I-0 & $0$ & $\frac{\epsilon}{16}$  & $0$ & $0$ & $0$ & 0 \\
I-1 & $0$ & $0$ & $\epsilon$ & $-\epsilon$ & $\frac{\epsilon}{2}$ & 2 \\
I-2 & $0$ & $\frac{\epsilon}{16}$ & $\epsilon$ & $\frac{16\epsilon}{3}$ & $-\frac{\epsilon}{2}$ & 1 \\
I-3 & $0$ & $0$ & $0$ & $\epsilon$ & $\frac{\epsilon}{4}$ & 3 \\
I-4 & $\epsilon$ & $0$ & $0$ & $0$ & $-\frac{\epsilon}{2}$ & 3 \\
I-5 & $0$ & $\frac{\epsilon}{16}$ & $0$ & $\epsilon$ & $-\frac{\epsilon}{8}$ & 0 \\
I-6 & $\epsilon$ & $\frac{\epsilon}{16}$ & $0$ & $0$ & $\frac{\epsilon}{4}$ & 0 \\
I-7 & $\frac{\epsilon}{3072}$ & $\frac{\epsilon}{32}$ & $\epsilon$ & $\frac{32 \epsilon}{3}$ & $-\frac{\epsilon}{8}$ & 2 \\
I-8 & $\frac{\epsilon}{3872}$ & $\frac{\epsilon}{22}$ & $\epsilon$ & $-\frac{88 \epsilon}{3}$ & $\frac{\epsilon}{22}$ & 1 \\
\end{tabular}
\end{ruledtabular}
\end{table}


Among eight fixed points, there are two most stable fixed points (label I-5 and I-6 in Table \ref{tab:table1}).
These scale invariant fixed points are stable up to fine-tuning more relevant terms $\Phi^2$ and $\Phi \partial_{\bar{\theta}} \partial_{{\theta}}  \Phi$. Note that the conformal fixed point discussed above is a special point $\tilde{\lambda}_{1} = 0$ and $\tilde{\lambda}_{-2} = 0$ not related by the field redefinition. In this sense, the scale invariance is more generic than the conformal invariance in this model.

We want to stress that when $\gamma \neq 0$ the (interacting) fixed point cannot be conformal even non-perturbatively because the two-point function $\langle \varphi(x) \omega(0) \rangle$ is non-zero but $\varphi$ and $\omega$ have the opposite anomalous dimensions $(\gamma, -\gamma)$, which is inconsistent with the conformal selection rules of the two-point function of primary operators. (The only loophole of this argument is when $\omega$ is a descendant of $\varphi$, which is not the case here perturbatively).

\section{Non-renormalization of the virial current}
What is the magic for this theory to avoid the genericity argument for the non-existence of the virial current?
To explain the mechanism, we observe that the candidate of the (non-conserved) virial current $V_\mu = \omega \partial_\mu \varphi - \varphi \partial_\mu \omega$ and a (conserved) supercurrent $q_\mu = {\Psi} \partial_\mu \omega - \omega \partial_\mu {\Psi}$ (and $\bar{q}_\mu = \bar{\Psi} \partial_\mu \omega - \omega \partial_\mu \bar{\Psi}$) are different component of the same superfield $\Phi \partial_\mu \partial_{\bar{\theta}}\partial_\theta \Phi - \partial_{\bar{\theta}}\partial_\theta \Phi \partial_\mu \Phi $. Then, we naturally expect that since the supercurrent is not renormalized, the virial current is not renormalized. 

The actual argument is a little more complicated because the dimension of the (conserved) supercurrent is {\it not } (necessarily) $d-1$. Without conformal symmetry and without superrotation symmetry, there is no simple way to determine the scaling weight of the supercoordinate $\theta$. Nevertheless, we show that the virial current has dimension $d-1$ exactly. Let $Q = \int d\Sigma_{\mu} q^\mu$ be the supertranslation charge, and $D$ be the scale charge. We have $[D,q_\mu] = \Delta_q q_\mu$, $[D,Q] = \Delta_Q Q = (\Delta_q - d+1)Q$. From the supertranslation, we have $[Q, V_\mu] = {q}_\mu$. Acting $D$, we obtain $\Delta_{Q} + \Delta_V = \Delta_q $, leading to the desired relation $\Delta_V = d-1$ even if $\Delta_Q \neq 0$.
Note that the argument here is non-perturbative.

One can show the same results by using correlation functions more explicitly.
Consider the Ward-Takahashi identity of the supertranslation:
\begin{align}
\langle \partial^\mu q_\mu(x_1) V_\nu (x_2) O(x_3) \rangle = \delta^{d} (x_1 - x_2) \langle {q}_\nu(x_2) O(x_3) \rangle,
\end{align}
where $O(x) = \partial_\theta Y(x,\theta)|_{\theta=0}$ is a certain $Q$ invariant operator that makes the equality non-vanish. (For the existence of such operators, we need $\lambda_{-1} \neq 0$. We can relax the condition $O(x)$ is $Q$ invariant at the expense of more terms on the right-hand side, in which case we focus on the scaling behavior at $x_1 = x_2$ .)
Equating the dimensions of both sides, we obtain $\Delta_q + 1 + \Delta_V + \Delta_O = d + \Delta_q + \Delta_O$, leading to the the desired relation $\Delta_V = d-1$ irrespective of $\Delta_q$.

One of the interesting features of the discussion here is that the divergence of the virial current $\partial^\mu (\omega \partial_\mu \varphi - \varphi \partial_\mu \omega)$ is non-zero, but it is supertranslation invariant (upon the use of the equations of motion). 
So there is no contradiction in that the non-conserved virial current is in the same multiplet with the conserved current.

Even without using the supertranslation invariance, we can offer an alternative argument as to why the virial current is not renormalized to six out of eight (label $1$ to $6$ in Table \ref{tab:table1}) scale invariant but not conformal fixed points. These six fixed points have $\lambda_1 = 0$ or $\lambda_{-1} = \lambda_{-2} = 0$, so all the coupling constants have non-positive (or non-negative) $SO(1,1)$ charge.\footnote{Fixed points I-1, I-3 
 and I-4 have an additional non-renormalization feature that they do not show any higher-loop corrections thanks to the spurious $SO(1,1)$ symmetry. The author would like to thank Slava Rychkov for pointing this out.}

How is this observation useful?
Let us suppose we want to compute the two-point function of the virial current $\langle V_\mu(x) V_\nu(0) \rangle $ perturbatively with respect to the coupling constants at hand. The crucial point is that due to the $SO(1,1)$ selection rule of the unperturbed theory,  $\langle V_\mu(x) V_\nu(0) \rangle $ will not depend on $\lambda_{-1}$ and $\lambda_{-2}$ at any orders of perturbation theory if $\lambda_1 = 0$ (this is simply because there are no positively charged operators under $SO(1,1)$ at hand). Now $\langle V_\mu(x) V_\nu(0) \rangle $ may depend on $\lambda_0$ but all the computations will reduce to the case without $\lambda_{-1}$ and $\lambda_{-2}$. When $\lambda_{-1} = \lambda_{-2}= 0$, however, the theory is $SO(1,1)$ symmetric, and $V_\mu$ is the conserved $SO(1,1)$ current, so it cannot acquire any anomalous dimensions. From the above discussions, this is true also when $\lambda_{-1} = \lambda_{-2} \neq 0$ as long as $\lambda_1 = 0$. The same argument holds when $\lambda_{-1} = \lambda_{-2} = 0$.

The scale but non-conformal fixed point with label $3$ has yet another argument based on the shift symmetry. The fixed point has $\lambda_0 = \lambda_1 = \lambda_{-1} = 0$, and $\varphi$  does not appear in the potential. Thus it has a shift symmetry under $\varphi \to \varphi + \text{const}$. The conserved shift current is $s_\mu = \partial_\mu \omega$ and the shift charge is $S = \int d \Sigma_\mu s^\mu$. Now, we may employ the argument given in \cite{Gimenez-Grau:2023lpz}. 

We start with $[D,s_\mu] = \Delta_s s_\mu$ or $[D,S] = (\Delta_s- (d-1)) S$. Acting $S$ on the virial current, we have $[S,V_\mu] = s_\mu$, and acting $D$, we obtain $\Delta_{s} + \Delta_V - (d-1) = \Delta_s$, leading to the desired relation $\Delta_V = d-1$ exactly.

We see that the argument based on the supercurrent and the argument based on the shift current are completely parallel. If we abstract the mechanism, what is needed is the symmetry $X$ that relates the virial current with the (conserved) current of the symmetry $X$. The existence of such $X$ guarantees that the dimension of the virial current is $d-1$ exactly.

\section{A comment on the original Parisi-Sourlas model without exra fine-tuning}
One may ask whether the fixed point of the original Paris-Sourlas model is conformal invariant and (even if so) whether we can find extra scale invariant but non-conformal fixed points. The original model was 
\begin{align}
S =& \int d^dx d^2\theta (\frac{-1}{2}\Phi(\partial^2_\mu - \partial_{\bar{\theta}} \partial_{{\theta}} )\Phi +\frac{\lambda_{1}}{4}\Phi^4 ) \cr
=& \int d^d x (-\omega \partial^2_\mu \varphi + \frac{\omega^2}{2} + \Psi \partial_\mu^2 \bar{\Psi} + \lambda_1 (\omega \varphi^3 -3 \Psi \bar{\Psi}\varphi^2)) \cr \label{original}
\end{align}
This section assumes some familiarity with the recent discussions (and terminology) on the Parisi-Sourlas supersymmetry in random field Ising model \cite{Kaviraj:2019tbg,Kaviraj:2020pwv,Kaviraj:2021qii}. We refer to \cite{Rychkov:2023rgq} for a pedagogical review.

Due to the superrotation $OSp(d|2)$, we find $x^\mu$, $\theta$ and $\bar{\theta}$ have the same scaling dimensions, so there is no renormalization group flow of the combination $-\varphi \partial_\mu^2 \omega +\frac{\omega^2}{2}$ and we find $\Delta_\omega - \Delta_\varphi = 2$ exactly. It means that we cannot perturbatively introduce non-zero anomalous dimension $\gamma$ that appeared in the candidate virial current but breaks the superrotation symmetry.
Thus we conclude the fixed point of the original Paris-Sourlas model is conformal invariant and we cannot find any other scale invariant but non-conformal fixed points, which can be checked explicitly within perturbation theory. This is in accord with the preservation of the conformal symmetry under the dimensional reduction $d\to d-2$. See e.g. \cite{Hoback:2020syd,Kaviraj:2022bvd,PhysRevE.108.044146,Cardy:2023zna,Trevisani:2024djr} for some recent discussions on dimensional reduction and Parisi-Sourlas supersymmetry.

Once we abandon the superrotation, things are less obvious. Indeed, the superrotation is just an accidental symmetry of the Parisi-Sourlas model from the random field theory viewpoint. While breaking of the superrotation symmetry is irrelevant near the upper-critical dimensions, it can be lost in lower dimensions \cite{Kaviraj:2020pwv}. The failure of the dimensional reduction as a consequence was observed in many works \cite{1984PhRvL..53.1747I,1987PhRvL..59.1829B,1998EL.....44...13B,2000cond.mat.10012F,Tarjus:2004wyx,Tissier:2011zz,Tarjus:2019zte,Balog:2020sre,2016PhRvL.116v7201F,2017PhRvE..95d2117F,Tarjus:2024mop}.
In the replica approach with the Cardy basis \cite{CARDY1985123} (where we have $n$ scalars reshuffled into $\omega$, $\varphi$ and $\chi_i$, $\chi_i$ being identified with fermions in the $n \to 0$ limit), the so-called SUSY writable leaders preserve supertranslation (but not superrotation). With only the supertranslation at hand, we may have a candidate for the virial current $V_\mu = \omega \partial_\mu\varphi - \varphi \partial_\mu \omega $, so possibly we expect scale invariance without conformal invariance.

It was further suggested that the supertranslation symmetry may be broken as well in lower dimensions. To analyze this effect, \cite{Kaviraj:2019tbg,Kaviraj:2020pwv,Kaviraj:2021qii} use the Cardy basis. In this basis, the breaking of the supertranslation is caused by the so-called SUSY non-writable leaders that become relevant. Without supertranslation symmetry, there is no sophisticated mechanism to protect the scaling dimension of the candidate virial current operator. The genericity argument, then, should lead to the conformal invariance of the random Ising model (without extra fine-tuning assumed in the other sections).

Is there any other supporting evidence of conformal invariance independent of the genericity argument?  
 One observation is that the candidate Virial current operator above $V_\mu = \omega \partial_\mu\varphi - \varphi \partial_\mu \omega$ cannot be represented in an $S_n$ invariant manner simply because the $S_n$ singlet vector operators constructed out of two fields and one derivative are necessarily total derivative of a $S_n$ singlet scalar operator and can be improved away (See e.g. \cite{1998EL.....44...13B,Kaviraj:2020pwv} for discussions on $S_n$ invariant operators). We recall that the non-existence of the tree-level candidate of the virial current was one of the bases of the conformal invariance of the critical Ising model in three dimensions, and the non-perturbative verification has been done in \cite{Delamotte:2015aaa,Meneses:2018xpu,Zhu:2022gjc}. Our evidence is still perturbative because with more fields one can construct the $S_n$ invariant vector operators that cannot be improved away.

Another potentially non-perturbative argument is to study the two-point function $\langle \omega \varphi \rangle$. While it is non-zero and cannot be consistent with conformal invariance if $\varphi$ and $\omega$ are both primaries, it can be consistent with the conformal invariance if the equations of motion suggest  $\omega \sim \partial^2_\mu \varphi$  and $\omega$ is a descendant of $\varphi$. The argument still relies on perturbation theory because the use of the equations of motion might be questioned in the non-perturbative regime.

Thus, in the non-perturbative regime, we eventually have to rely on the genericity argument, which is based on the claim that there is no other ``sophisticated mechanism" to protect the scaling dimensions of virial current without supertranslation symmetry. To see the validity of the genericity argument in a broader context, we now come back to the fine-tuned model with only quartic potential and look for non-supersymmetric scale invariant but not conformal fixed points. The existence may force us to reconsider the genericity argument.


\section{Fixed points without supertranslation symmetry}\label{general}
We have seen supertranslation symmetry plays a significant role in obtaining scale invariance without conformal invariance in our model. What happens if we abandon the supertranslation symmetry and consider more generic quartic interactions among $\varphi$,$\omega$,$\Psi$, and $\bar{\Psi}$? The action we study is 
\begin{align}
S_B = &\int d^d x (- \omega \partial_\mu^2 \varphi + {\Psi} \partial_\mu^2 \bar{\Psi}  \cr
&+\lambda_{2} \varphi^4 + \lambda_{1} \omega\varphi^3+ \lambda_0 \varphi^2 \omega^2  + \lambda_{-1} \varphi \omega^3 + \lambda_{-2} \omega^4 \cr &+ y_{1} \Psi \bar{\Psi} \varphi^2 + y_0 \Psi \bar{\Psi} \varphi \omega + y_{-1} \Psi \bar{\Psi} \omega^2) .  
\end{align}

By computing the one-loop beta functions in $d=4-\epsilon$ dimensions, we obtain the scale invariant condition as {\small
\begin{align}
\epsilon \lambda_2 + 4\gamma \lambda_2 &= 24 \lambda_0 \lambda_2 +9 \lambda_1^2 -y_1^2 \cr
\epsilon \lambda_1 + 2 \gamma \lambda_1 & = 72 \lambda_2 \lambda_{-1} + 36 \lambda_1 \lambda_0 -2y_1 y_0  \cr
\epsilon \lambda_0 &= 144\lambda_2 \lambda_{-2} +54 \lambda_1 \lambda_{-1} + 20 \lambda_0^2 -2 y_{1} y_{-1} - y_0^2  \cr
\epsilon \lambda_{-1} -2 \gamma \lambda_{-1} & = 72 \lambda_{-2} \lambda_{1} + 36 \lambda_{-1} \lambda_0 -2y_{-1} y_0  \cr
\epsilon \lambda_{-2} - 4\gamma \lambda_{-2} &= 24 \lambda_0 \lambda_{-2} +9 \lambda_{-1}^2 -y_{-1}^2 \cr
\epsilon y_1 + 2\gamma y_1 &=  -8 y_1 y_0 +4\lambda_0 y_1 +6 y_0 \lambda_1 +24 y_{-1}\lambda_2 \cr
\epsilon y_0  &= -4 y_0^2 -16 y_1 y_{-1} +12 y_1 \lambda_{-1} + 8 y_0 \lambda_0 +12y_{-1}\lambda_1  \cr
\epsilon y_{-1} - 2\gamma y_{-1} &=  -8 y_{-1} y_0 +4\lambda_0 y_{-1} +6 y_0 \lambda_{-1} +24 y_{1}\lambda_{-2}
\end{align}}
As before the fixed point with $\gamma=0$ is conformal invariant and otherwise it is only scale invariant. If we set $\lambda_2 = 0$, $y_1= -3 \lambda_1 $, $y_0 = -2\lambda_0$ and $y_{-1} = -\lambda_{-1}$, we recover supertranslation symmetry.

There are numerous non-supersymmetric fixed points. In particular, we find eleven additional conformal fixed points with real coupling constants as shown in Table \ref{tab:table2}. (Conformal fixed points with imaginary coupling constants also exist.)

\begin{table}[b]%
\caption{\label{tab:table2}%
Non-trivial conformal fixed points we found. We have fixed the $SO(1,1)$ field redefinition ambiguity by $\lambda_2 = \epsilon$ or $y_1 = \epsilon$. 
The fixed points  II-10 and II-11 have too lengthy analytical expressions to be listed here.  Numerically, ({$\lambda_2 = \epsilon$, $\lambda_1 = 0.38916 \epsilon$, $\lambda_0 = -0.00538145\epsilon$, $\lambda_{-1} = 0.00287702 \epsilon$,  
 $\lambda_{-2} = 0.000054655 \epsilon$, $y_1 = 0.483586 \epsilon$, $y_0 = -0.266144\epsilon$ , $y_{-1} = 0.0035751\epsilon$}), and ($\lambda_2 = \epsilon$, $  \lambda_1 = 0.089301 \epsilon$,  $\lambda_0 = 0.0561061 \epsilon$, $\lambda_{-1}= 0.000202618 \epsilon$, $\lambda_{-2} = 5.14807\times10^{-6} \epsilon$, 
$ y_1 = -0.646775 \epsilon$, $y_0 = -0.0816817 \epsilon$, $y_{-1} = -0.00146749 \epsilon$).},

\begin{ruledtabular}
\begin{tabular}{lcccccccc}
\textrm{Label}&
$\lambda_2$&
$\lambda_1$ &
$\lambda_{0}$&
$\lambda_{-1}$&
$\lambda_{-2}$&
$y_{1}$&
$y_{0}$&
$y_{-1}$
\cr
\colrule
II-1 & $0$ & $\frac{\epsilon}{3}$  & $0$ & $\frac{\epsilon}{432}$ & $0$ & $\epsilon$ & $-\frac{\epsilon}{6}$ & $\frac{\epsilon}{144}$ \\
II-2 & $0$ & $0$  & $-\frac{\epsilon}{16}$ & $0$ & $0$ & $0$ & $-\frac{3\epsilon}{8}$ & $0$ \\
II-3 & $0$ & $0$  & $\frac{\epsilon}{20}$ & $0$ & $0$ & $0$ & $0$ & $0$ \\
II-4 & $\epsilon$ & $-\frac{\epsilon}{2 \sqrt{6}}$  & $\frac{\epsilon}{24}$ & $\frac{\epsilon}{72 \sqrt{6}}$ & $\frac{\epsilon}{1296}$ & $-\frac{\sqrt{6}\epsilon}{4}$ & $-\frac{\epsilon}{4}$ & $\frac{\epsilon}{24 \sqrt{6}}$  \\
II-5 & $\epsilon$ & $0$  & $\frac{\epsilon}{24}$ & $0$ & $\frac{5\epsilon}{20736}$ & $0$ & $-\frac{\epsilon}{6}$ & $0$  \\
II-6 & $\epsilon$ & $\frac{\epsilon}{\sqrt{14}}$  & $\frac{3\epsilon}{112}$ & $\frac{\epsilon}{224 \sqrt{14}}$ & $\frac{\epsilon}{50176}$ & $-\frac{\sqrt{14}{}\epsilon }{7}$ & $-\frac{\epsilon}{14}$ & $-\frac{\epsilon}{112 \sqrt{14}}$  \\
II-7 & $\epsilon$ & $\frac{\epsilon}{6 \sqrt{14}}$  & $\frac{\epsilon}{21}$ & $\frac{\epsilon}{756 \sqrt{14}}$ & $\frac{\epsilon}{15876}$ & $-\frac{3 \epsilon}{2 \sqrt{14}} $ & $-\frac{\epsilon}{14}$ & $-\frac{\epsilon}{84 \sqrt{14}}$  \\
II-8 & $\epsilon$ & $\frac{\epsilon}{3 \sqrt{2}}$  & $\frac{\epsilon}{48}$ & $\frac{\epsilon}{864 \sqrt{2}}$ & $\frac{\epsilon}{82944}$ & $0$ & $0$ & $0$  \\
II-9 & $\epsilon$ & $0$  & $\frac{\epsilon}{24}$ & $0$ & $\frac{\epsilon}{20736}$ & $0$ & $0$ & $0$  \\
II-10 & $\epsilon$ & $*$  & $*$ & $*$ & $*$& $ * $ & $*$ & $*$ \\
II-11 & $\epsilon$ & $*$  & $*$ & $*$ & $*$& $ * $ & $*$ & $*$ \\

\end{tabular}
\end{ruledtabular}
\end{table}

Let us now move on to the scale-invariant but not conformal fixed points.
For instance, let us study the fixed point with $\lambda_{2} = 0$. Supertranslation invariance demands this condition, but there are other solutions without supertranslation symmetry. In addition to the three conformal fixed points already mentioned above, we find eight scale invariant but not conformal fixed points without supertranslation and one fixed line (i.e. fixed points with a moduli). They are listed in table \ref{tab:table3}.

\begin{table}[b]%
\caption{\label{tab:table3}%
Scale invariant fixed points with $\lambda_2 = 0$. We have fixed the $SO(1,1)$ field redefinition by demanding $y_{-1} = \epsilon$ (or $\lambda_{-1} = \epsilon $ or $\lambda_{-2} = \epsilon$  or $\lambda_1 = \epsilon$ when it is zero). A free parameter $\tilde{\lambda}$ in fixed point 4 is a moduli.},

\begin{ruledtabular}
\begin{tabular}{lcccccccc}
\textrm{Label}&
$\lambda_1$ &
$\lambda_{0}$&
$\lambda_{-1}$&
$\lambda_{-2}$&
$y_{1}$&
$y_{0}$&
$y_{-1}$&
$\gamma$
\cr
\colrule
III-1 & $0$ & $0$  & $\epsilon$ & $-9\epsilon$ & $0$ & $0$ & $0$ & $\frac{\epsilon}{2}$ \\
III-2 & $0$ & $\frac{\epsilon}{20}$  & $\epsilon$ & $\frac{45\epsilon}{7}$ & $0$ & $0$ & $0$ & $-\frac{2\epsilon}{5}$ \\
III-3 & $0$ & $\frac{\epsilon}{20}$  & $0$ & $\epsilon$ & $0$ & $0$ & $0$ & $-\frac{\epsilon}{20}$ \\
III-4 & $0$ & $0$  & $\tilde{\lambda}$ & $\epsilon - \frac{9 \tilde{\lambda}^2}{\epsilon}$ & $0$ & $0$ & $\epsilon$ & $\frac{\epsilon}{2}$ \\
III-5& $0$ & $\frac{\epsilon}{16}$  & $-\frac{\epsilon}{3}$ & $0$ & $0$ & $-\frac{\epsilon}{8}$ & $\epsilon$ & $-\frac{\epsilon}{4}$ \\
III-6& $0$ & $\frac{\epsilon}{20}$  & $0$ & $\frac{5}{9}\epsilon$ & $0$ & $0$ & $\epsilon$ & $\frac{2\epsilon}{5}$ \\
III-7& $0$ & $-\frac{\epsilon}{16}$  & {\scriptsize $\frac{10-\sqrt{73}}{9}\epsilon$} & {\scriptsize $\frac{8 \left(17 \sqrt{73}-149\right)}{207} \epsilon$ } & $0$ & $-\frac{3\epsilon}{8}$ &$\epsilon$ & {\scriptsize $\frac{3-\sqrt{73}}{8} \epsilon$} \\
III-8& $0$ & $-\frac{\epsilon}{16}$  & {\scriptsize $\frac{10+\sqrt{73}}{9}\epsilon$} & {\scriptsize $\frac{8 \left(17 \sqrt{73}+149\right)}{207} \epsilon$ } & $0$ & $-\frac{3\epsilon}{8}$ &$\epsilon$ & {\scriptsize $\frac{3+\sqrt{73}}{8} \epsilon$} \\
III-9 & $ 0 $ & $ -\frac{\epsilon}{16}$  & $0$ & $\epsilon$ & $0$ & $-\frac{3\epsilon}{8} $ &$0$ & $\frac{5\epsilon}{8}$ \\

\end{tabular}
\end{ruledtabular}
\end{table}

What is the sophisticated mechanism to protect the scaling dimension of the virial current without supertranslation symmetry? We find that all the fixed points in table \ref{tab:table3} have either $\lambda_1 = y_1 = 0$ or $\lambda_{-1}=\lambda_{-2} = y_{-1} = 0$, so as discussed in section \ref{search}, the two-point function of the virial current only depends on $\lambda_0$ and $y_0$ and the scaling dimension is protected. 

Some of the fixed points without supertranslation symmetry have $y_1 = y_0 = y_{-1} = 0 $, and the fermions $\Psi$ and $\bar{\Psi}$ are decoupled. In the decoupled bosonic sector, we have only $\varphi$ and $\omega$, and they define the simplest interacting field theory with $SO(1,1)$ invariant kinetic term. 

If we look at the bosonic decoupled sector of fixed points III-1,III-2,III-3 in Table \ref{tab:table3}, we find something interesting. These scale invariant but not conformal fixed points have only $\mathbb{Z}_2$ symmetry $(\varphi,\omega) \to (-\varphi,-\omega)$ and there is no other global symmetry. (Fixed point III-3 has a mock $\mathbb{Z}_2$ symmetry $(\varphi, \omega) \to (i\varphi, -i\omega)$, which should not be regarded as a symmetry in the usual sense because the transformation violates the reality condition.) These are (non-unitary) ``counterexamples" of the conjecture that there does not exist any non-trivial fixed point with only one $\mathbb{Z}_2$ global symmetry within one-loop multi-scalar fixed points with only one conserved stress tensor in $d=4-\epsilon$ dimensions \cite{Rychkov:2018vya}.

So far, we have given the ``sophisticated argument" why the virial current is not normalized in scale-invariant but not conformal fixed points. Interestingly, if we relax the condition $\lambda_2 = 0$, we have found two mysterious scale invariant but not conformal fixed points with $y_1 = y_0 = y_{-1} = 0$.

If we fix the $SO(1,1)$ field definition ambiguity by $\lambda_{1}= \epsilon$, they are located at $\lambda_2 = \frac{1}{3} \left(\mp 16 \sqrt{10}-20\right)\epsilon $, $\lambda_1 = \epsilon$, $\lambda_0 =  \frac{1}{32} \epsilon $, $\lambda_{-1}= \frac{1}{3840} \epsilon$, $\lambda_{-2} = \frac{\pm 4 \sqrt{\frac{2}{5}}-1}{2211840} \epsilon$ and $\gamma = \mp\frac{1}{2 \sqrt{10}}\epsilon$. They do not have supertranslation symmetry, nor can we employ the ``triangular structure", which applies when we have coupling constants only with non-negative or non-positive $SO(1,1)$ charge, to show the non-renormalization of the virial current.

We have not found any good non-accidental reason why the virial current is not renormalized in these mysterious fixed points. We leave the enigmas for future study.

\section{Discussions and Conclusion}
In this paper, inspired by the possibility of emergent supersymmetry in fine-tuned random systems, we have studied a field theory model with a quartic potential of one superfield, possessing the Parisi-Sourlas supertranslation symmetry. Within perturbative $\epsilon$ expansion, we have found eight scale invariant but not conformal renormalization group fixed points. The scaling dimension of the virial current is protected thanks to the supertranslation symmetry.

Once the supertranslation symmetry is broken, as in the random field Ising model without extra fine-tuning, we currently do not understand why the virial current is not renormalized. From the genericity argument, we propose that the random field Ising model without extra fine-tuning would show conformal invariance, and this is probably a commonly believed scenario (see e.g. conformal bootstrap study in \cite{Hikami:2018mrf}, where conformal invariance is assumed). Still, it is worthwhile checking the conformal symmetry explicitly in simulations or experiments. We may find a surprise.

\begin{acknowledgments}
YN is in part supported by JSPS KAKENHI Grant Number 21K03581. The author would like to thank Slava Rychkov and Andreas Stergiou for the correspondence.
\end{acknowledgments}

\bibliography{apssamp}

\appendix

\section{Corresponing $PT$ symmetric quantum field theory with $O(2)$ invariant kinetic term}
In the main text, we studied Euclidean statistical field theories, and in this Appendix, we propose the corresponding quantum field theories with the more conventional $O(2)$ invariant kinetic term.
Let us focus on the decoupled bosonic sector of the action in section VI and do the field redefinition $\omega = \phi_1 + i\phi_2$ and $\varphi = \phi_1 - i \phi_2$. We obtain the analytically-continued action
\begin{align}
S = \int d^dx &(\partial_\mu \phi_1 \partial_\mu \phi_1 + \partial_\mu \phi_2 \partial_\mu \phi_2) \cr
&+ \lambda_2 (\phi_1-i\phi_2)^4 + \lambda_1 (\phi_1-i\phi_2)^2(\phi_1^2+\phi_2^2) \cr
& + \lambda_0 (\phi_1^2 + \phi_2^2)^2   + \lambda_{-1} (\phi_1+i\phi_2)^2(\phi_1^2 + \phi_2^2) \cr
&+ \lambda_{-2}(\phi_1 + i\phi_2)^4 \ .
\end{align}
If we treat $\phi_1$ and $\phi_2$ as real scalar fields, we have $O(2)$ invariant kinetic term rather than $O(1,1)$ invariant kinetic term. If we wish, we may further rotate the time direction $x_d \to i x_0$ to obtain the action for a quantum field theory in the Lorentzian signature.

The renormalization group beta function within the $\epsilon$ expansion is exactly the same as in section VI (with $y_0 = y_1 = -y_{-1}=0$) and we find three non-trivial conformal fixed points (corresponding to II-3, II-8 and II-9) and three scale invariant but not conformal fixed points (corresponding to III-1, III-2 and III-3). More precisely, they come with a one-parameter family because the $O(1,1)$ field redefinition cannot be done here.

As we discussed in section VI, the scale-invariant but not conformal fixed points corresponding to III-1 and III-2 are of particular interest because the fixed point has only one global $\mathbb{Z}_2$ symmetry $(\phi_1, \phi_2) \to (-\phi_1,- \phi_2)$. The fixed point III-3 has an additional exchange symmetry $(\phi_1, \phi_2) \to (\phi_2, \phi_1)$, which was a mock symmetry in $\omega$-$\varphi$ basis.  These fixed points are not real but $PT$ symmetric i.e. invariant under the $PT$ conjugation: $i \to -i$ and $\phi_2 \to -\phi_2$. All the real fixed points (i.e. $\lambda_2 = \lambda_{-2}$ and $\lambda_1 = \lambda_{-1}$) are conformal invariant as expected from the analysis by Polchinski, and they are equivalent to the Wilson-Fisher $O(2)$ fixed point, the two decoupled Ising fixed points, and the decoupled Ising fixed point and Gaussian fixed point.

\section{Ambiguity in beta functions}
The general form of the trace of the stress tensor
\begin{align}
T = \beta_a O^a + \partial_\mu J_\mu + \partial_\mu \partial_\nu L_{\mu\nu}
\end{align}
suggests that the beta function $\beta_a$ may be ambiguous when we can use the equations of motion to rewrite $\partial_\mu J^\mu$ into a sum of non-derivative operators. For our purpose of discussing the scale invariance without conformal invariance, it is important to compute not only $\beta_a$ but the ambiguity-free part of $\beta_a O^a + \partial_\mu J_\mu$, which is sometimes referred to as $B_aO^a$ in the literature and see whether it results in a non-zero virial current.  The $L_{\mu\nu}$ term is not as important because they can be improved away. In the standard flat space perturbation theory, such analysis could be non-trivial because we might miss the total derivative terms in the (global) renormalization group transformation. 

In our case, the potentially relevant contribution comes from $J_\mu = f(\lambda,y) (\omega \partial _\mu \varphi - \varphi \partial_\mu \omega)$.
We here argue that with the prescription used in the main text (e.g. the general model presented in section \ref{general} with the one-loop beta function presented there), there is no additional contribution from $\partial_\mu J_\mu$ term within the one-loop order we are interested in, so we do not have to worry about potential cancellation from the ambiguity of the beta functions (computed with the prescription used in the main text).  

To constrain the total derivative ambiguities in the renormalization group to some extent, we use the prescription that the renormalization group flow preserves the spurious $SO(1,1)$ symmetry and the exchange of $\varphi$ and $\omega$. Spurious here means we also transform the coupling constants under the $SO(1,1)$ symmetry and the exchange symmetry. More explicitly, the former induces  $(\lambda_{\pm1},\lambda_{\pm 2},y_{\pm 1}) \to (e^{\pm 2 \rho}\lambda_{\pm1}, e^{\pm 4\rho} \lambda_{\pm 2}, e^{\pm 2 \rho} y_{\pm 1}) $, and the latter induces $(\lambda_{\pm 1}, \lambda_{\pm 2}, y_{\pm 1}) \to (\lambda_{\mp 1}, \lambda_{\mp 2}, y_{\mp 1})$. We then assume that $J_\mu$ is invariant under the spurious symmetry transformation. It is crucial to note that our beta functions presented in the main text satisfy this condition.

We can easily see that at the linear order in the coupling constants, without doing any Feynman diagram computation, there is no candidate for $J_{\mu}$. (Independently, we recall that there is no wave-function renormalization in the $\phi^4$-like theories with the conventional one-loop Feynman diagram computation even in a non-uniform background, so this term is indeed absent).
The first non-trivial term might appear at two-loop with $f(\lambda,y) \propto (\lambda_{1} y_{-1} -  \lambda_{-1} y_1)$, which is equivalent to $O(\lambda^3,  \lambda^2 y , \lambda y^2)$ in beta functions. Thus, we can conclude that the potential ambiguity that may affect the trace of the stress tensor cannot alter the one-loop results in the main text (but it might at the higher order: actually, to obtain a non-zero diagram, we need at least four $\Psi$, so this two-loop term should be absent in the actual computation). When $y=0$, the first non-trivial contribution may come from $f(\lambda) \propto (\lambda_{2} \lambda_{-1}^2 - \lambda_{-2} \lambda_1^2 )$, which is even smaller. We have not studied whether this term appears in the actual computation, but, if any, it cannot affect our one-loop conclusion.

\end{document}